\documentclass[%
 aps,
 prl,% Physical Review Letters
 preprint,% Change to reprint for final journal formatting
 floatfix,% Prevent stuck floats
 groupedaddress,
 amsmath,amssymb,
 superscriptaddress,
]{revtex4-2}
\usepackage{graphicx}
\usepackage{dcolumn}
\usepackage{bm}
\usepackage{hyperref}

\begin{document}

\title{Intermittency-Driven Turbulence Cascade Memory Extends the Markov--Einstein Coherence Length Beyond the Canonical Estimate}

\author{Y.~Sungtaek~Ju}
\email{sungtaek.ju@ucla.edu}
\affiliation{Department of Mechanical and Aerospace Engineering,
University of California, Los Angeles, California 90095, USA}

\date{\today}

\begin{abstract}
Using direct numerical simulation of forced isotropic turbulence at $\text{Re}_\lambda \approx 1300$ and $\approx 433$, together with two independent Markov-by-construction null surrogates, we measure the Markov--Einstein coherence length of the turbulent energy cascade to be $\Delta r \approx 3.2$--$3.6$ in log-scale cascade coordinates---approximately three times the canonical estimate $\Delta r \approx 1$.  Stratifying the gap-scan test by local dissipation intensity and by increment amplitude reveals that intermittent events carry $\Delta r \approx 3$--$4$, while at mid-inertial-range scales the quiescent cascade recovers $\Delta r \approx 1.0$--$1.4$, consistent with the canonical value.  Near the dissipation range this pattern reverses: bulk dynamics carry more memory than extreme events, consistent with the spectral bottleneck.  The excess memory is internal to the inertial range and Reynolds-number-independent over $\text{Re}_\lambda \approx 433$--$1300$.  These findings indicate that the Markov approximation underlying the cascade Fokker--Planck equation and fluctuation-theorem analyses is substantially more restrictive than previously assumed, and that a non-Markovian correction, informed by the amplitude-dependent memory structure identified here, is needed for the intermittent component of the cascade.
\end{abstract}

\maketitle

\emph{Introduction.}---The energy cascade in fully developed turbulence transfers kinetic energy from large to small scales through a hierarchy of nonlinear interactions. Friedrich and Peinke~\cite{Friedrich1997} showed that the conditional probability density of longitudinal velocity increments $\delta u_\ell = [\bm{u}(\bm{x}+\ell\hat{\bm{e}}) - \bm{u}(\bm{x})]\cdot\hat{\bm{e}}$ across scales obeys a Fokker--Planck equation in the cascade coordinate $r = -\ln(\ell/L)$, with drift and diffusion coefficients estimable directly from data.  This framework, reviewed comprehensively in Ref.~\cite{Friedrich2011}, rests on the Markov property: the conditional statistics of $\delta u$ at scale $r$ depend only on $\delta u$ at the immediately preceding scale, not on the full history.

The Markov assumption enables a powerful factorization of the multi-scale joint PDF into products of transition probabilities~\cite{Renner2001}, reducing the full complexity of the cascade to a single Fokker--Planck equation.  This has been exploited to establish universality of small-scale statistics across different turbulent flows~\cite{Renner2002,Stresing2010}, to connect the cascade to non-equilibrium thermodynamics via the integral fluctuation theorem (IFT)~\cite{Nickelsen2013,Reinke2018,Fuchs2020}, and to construct kinetic Fokker--Planck closure models for turbulence~\cite{Cao2025}.  All of these applications assume the Markov property holds at scale separations $\Delta r \gtrsim 1$.

The validity of this assumption was tested by Renner, Peinke, and Friedrich~\cite{Renner2001} using cryogenic helium jet data at $\text{Re}_\lambda \sim 200$--$700$.  They applied the Wilcoxon rank-sum test to probe conditional independence and found the cascade to be Markovian for separations exceeding a coherence length~$\lambda_\mathrm{EM}$. L\"uck et al.~\cite{Lueck2006} formalized the gap-scan procedure and connected $\lambda_\mathrm{EM}$ to the Taylor microscale. Stresing et al.~\cite{Stresing2011} consolidated the measurement across multiple flows (free jets, grid turbulence, cylinder wakes) and nesting structures, reporting $\lambda_\mathrm{EM} \approx (0.8 \pm 0.2)\lambda_T$ in physical units, which corresponds to $\Delta r \approx 1$ in the cascade coordinate (see End Matter for the conversion).  This estimate has been reproduced across a range of experimental configurations.

Two developments motivate a re-examination.  First, Sinhuber, Bewley, and Bodenschatz~\cite{Sinhuber2017} discovered oscillations in velocity structure functions within the inertial range at high Reynolds numbers ($\text{Re}_\lambda$ up to 1600), demonstrating that dissipation influences inertial-range statistics at scales significantly larger than predicted.  If dissipation-range dynamics contaminate the conditional structure of the cascade, they could inflate the apparent coherence length.  Second, the original experimental measurements relied on hot-wire anemometry with the Taylor frozen-turbulence hypothesis, ${\sim}10^5$ samples per scale, and no independent Markov-by-construction null surrogates to calibrate the Wilcoxon test's rejection floor.  Modern DNS databases provide $10^{6+}$ samples without probe artifacts, and surrogate construction is straightforward.

In this Letter, we present four results: (i)~a revised measurement $\Delta r \approx 3.2$--$3.6$ at $\text{Re}_\lambda \approx 1300$; (ii)~evidence that intermittent events drive the excess memory while, at the cleanest inertial-range center, the quiescent cascade recovers the canonical $\Delta r \approx 1$; (iii)~a mechanism reversal near the dissipation range; and (iv)~Reynolds-number independence of the finding at $\text{Re}_\lambda \approx 433$.

\emph{Data and methods.}---We use two forced homogeneous isotropic turbulence datasets from the Johns Hopkins Turbulence Databases~\cite{Li2008}.  The primary dataset (isotropic8192, $\text{Re}_\lambda \approx 1300$~\cite{Yeung2015}) is an $8192^3$ DNS with $\nu = 4.385 \times 10^{-5}$, $\eta \approx 5.0 \times 10^{-4}$, and $\lambda_T \approx 0.018$.  Ten $512^3$ velocity cubes are extracted from one snapshot via SciServer's container-local Zarr backend.  Longitudinal velocity increments $\delta u_\ell$ are computed on a 30-scale log-spaced grid from $\ell_\mathrm{min} = 8\eta$ to $\ell_\mathrm{max} = 0.385$ (step $\Delta r_\mathrm{step} = 0.157$) using an aligned-samples protocol: for each cube and axis direction, spatial positions are drawn once and reused across all scales, so that sample~$j$ at scale~$\ell_1$ and sample~$j$ at scale~$\ell_2$ correspond to the same spatial realization.  This alignment is essential. Independent per-scale subsampling would trivialize the conditional-independence structure the Markov test probes.  Approximately $10^6$ samples per scale are pooled across cubes and axes.

The cross-check dataset (isotropic1024coarse, $\text{Re}_\lambda \approx 433$~\cite{Li2008}) is a $1024^3$ DNS with $\nu = 1.85 \times 10^{-4}$, $\eta \approx 2.87 \times 10^{-3}$, and $\lambda_T \approx 0.118$.  Ten $256^3$ cubes, 15-scale grid from $8\eta$ to 0.35 ($\Delta r_\mathrm{step} = 0.195$).  The narrower inertial range at this Re limits the maximum scan reach to $\Delta r_\mathrm{max} = 2.72$. Any $\Delta r$ exceeding this is reported as a lower bound.

The Markov property can be tested either via the Chapman--Kolmogorov~(CK) equation for transition probabilities~\cite{Renner2001} or via conditional-independence tests.  We use the Wilcoxon rank-sum gap-scan~\cite{Renner2001,Lueck2006,Stresing2011} because it provides a scalar test statistic amenable to gap-scan analysis and stratification by local flow properties, which is essential for our central finding.  A CK verification is presented in the Supplemental Material~\cite{SM}.  For a scale triple $(\ell_1 < \ell_2 < \ell_3)$ with indices $(i_1, i_2, i_3)$, $\delta u_{\ell_2}$ is quantile-binned into $n_2 = 8$ bins.  Within each bin, $\delta u_{\ell_3}$ is sub-binned into $n_3 = 8$ bins.  The $\delta u_{\ell_1}$ distributions are compared across all $\binom{n_3}{2} = 28$ sub-bin pairs via two-sided Mann--Whitney~$U$~\cite{Mann1947}.  Under the Markov property $p(\delta u_{\ell_1} \mid \delta u_{\ell_2}, \delta u_{\ell_3}) = p(\delta u_{\ell_1} \mid \delta u_{\ell_2})$, conditioning on $\delta u_{\ell_2}$ renders $\delta u_{\ell_1}$ independent of $\delta u_{\ell_3}$, and the rejection fraction should equal $\alpha = 0.05$.  Substantially higher fractions indicate non-Markov memory.

The gap scan varies the total separation $s = i_3 - i_1$ at fixed center~$i_2$.  We extend the classical symmetric parameterization~\cite{Renner2001} to asymmetric triples, going asymmetric only when the symmetric triple would exceed the grid boundary.  Per-sub-bin samples are capped at 300 to control the test's unbounded power at large~$N$, following Stresing et al.~\cite{Stresing2011}.  The qualitative finding that $\Delta r$ substantially exceeding the canonical value is robust to binning parameters: varying $(n_2, n_3)$ over $\{6,8,10\}^2$ and max\_per\_bin over $\{200,300,500\}$, all resolved combinations give $\Delta r \geq 2.2$ (see Supplemental Material~\cite{SM}).  The specific threshold~$s^*$ varies with the statistical power controlled by max\_per\_bin; we report results at the baseline ($n_2 = n_3 = 8$, max\_per\_bin $= 300$).  The coherence length is $\Delta r = s^* \times \Delta r_\mathrm{step}$, where $s^*$ is the smallest separation at which the DNS excess over the surrogate baseline drops to $\leq 0.05$ and stays below at all larger separations.

Two Markov-by-construction surrogates pin the rejection-fraction floor.  (i)~A shuffle surrogate: per-scale independent permutation of the aligned DNS increments, preserving marginal PDFs exactly while destroying all cross-scale correlation.  (ii)~A DNS-matched Ornstein--Uhlenbeck chain on the DNS $r$-grid with matched per-scale variance (noise fraction $f = 0.5$).  Both return flat rejection fractions at $\alpha \approx 0.05$ across all separations, confirming the test is well-calibrated.  Details of surrogate construction are in the Supplemental Material~\cite{SM}.

Two stratification analyses are applied at each center.  First, dissipation conditioning: samples are split by the median of a local dissipation proxy $\varepsilon_\ell = |\delta u_{\ell_2}|^3/\ell_2$ into quiescent (low-$\varepsilon$) and intermittent (high-$\varepsilon$) strata.  Second, increment-amplitude stratification: samples are split by the 10th and 90th percentiles of $\delta u_{\ell_2}$ into a central 80\% (core) and an outer 20\% (tails).  Both stratifications use the same gap-scan procedure on the masked subsets, with the shuffle surrogate identically masked as the null baseline.

\begin{figure}[htb]
\includegraphics[width=0.7\columnwidth]{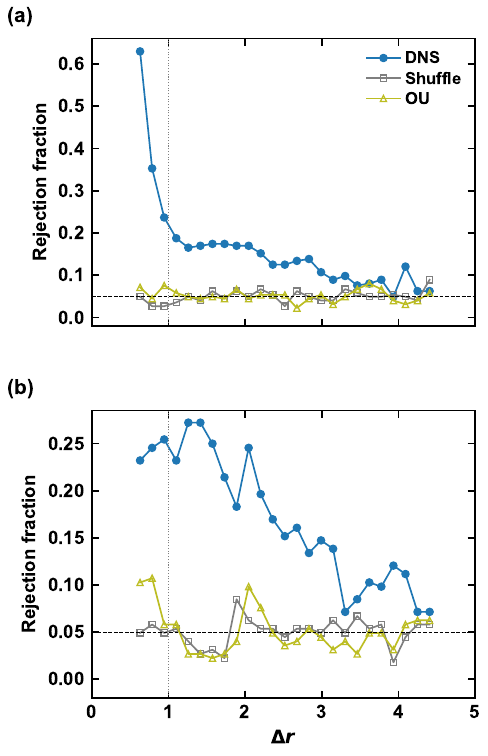}
\caption{Gap-scan at $\text{Re}_\lambda \approx 1300$.  Rejection fraction vs.\ total separation $\Delta r$ for centers (a)~$c = 11$ ($\ell_2/\lambda_T = 1.26$) and (b)~$c = 14$ ($\ell_2/\lambda_T = 2.02$).  Filled circles: DNS.  Open squares: shuffle surrogate.  Open triangles: OU surrogate.  Dashed line: nominal $\alpha = 0.05$.  Dotted vertical line: canonical $\Delta r = 1$~\cite{Stresing2011}.}
\label{fig:gap_scan}
\end{figure}

\emph{Results.}---Figure~\ref{fig:gap_scan} shows the gap-scan curves at $\text{Re}_\lambda \approx 1300$ for the two cleanest inertial-range centers.  The DNS rejection fraction remains well above the surrogate baselines out to $s \approx 20$ ($\Delta r \approx 3.1$) before dropping to the floor.  At center $c = 11$ ($\ell_2/\lambda_T = 1.26$), $s^* = 20$ gives $\Delta r = 3.15$.  At $c = 14$ ($\ell_2/\lambda_T = 2.02$), $s^* = 21$ gives $\Delta r = 3.31$.  At $c = 17$ ($\ell_2/\lambda_T = 3.23$), $s^* = 23$ gives $\Delta r = 3.62$.  The measured $\Delta r$ is approximately constant across the inertial range and exceeds the canonical $\Delta r \approx 1$ by a factor of~3.

At $c = 8$ ($\ell_2 = 0.78\lambda_T$, near the dissipation range), $\Delta r = 4.09$, which is elevated relative to the inertial-range values, consistent with dissipation-range contamination~\cite{Sinhuber2017}.

\begin{figure*}[htb]
\includegraphics[width=\textwidth]{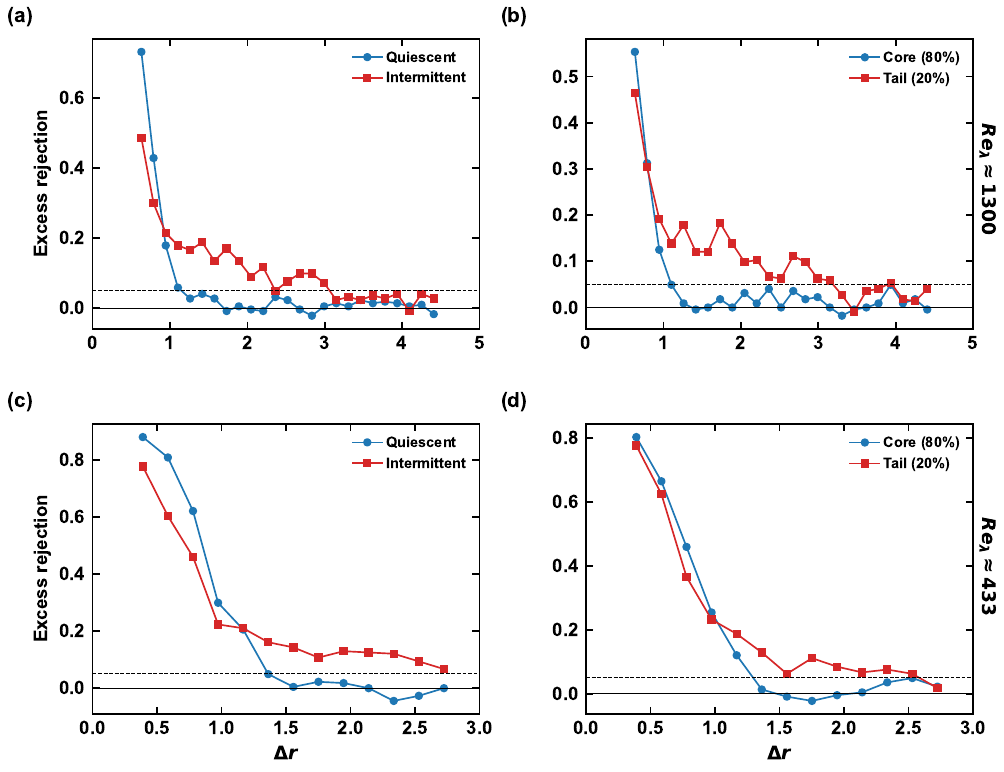}
\caption{Stratified excess (DNS $-$ shuffle rejection fraction) at both Re.  Top row: $\text{Re}_\lambda \approx 1300$, $c = 11$ ($\ell_2/\lambda_T = 1.26$).  Bottom row: $\text{Re}_\lambda \approx 433$, $c = 8$ ($\ell_2/\lambda_T = 0.93$).  Left column: dissipation conditioning (blue: quiescent, red: intermittent).  Right column: increment-amplitude stratification (blue: central 80\%, red: outer 20\%).  Horizontal dashed line: threshold at 0.05.  At these centers, the quiescent/core stratum drops to the Markov floor at $\Delta r \approx 1$ while the intermittent/tail stratum remains elevated.}
\label{fig:stratified}
\end{figure*}

Figure~\ref{fig:stratified} shows the stratified excess curves (DNS $-$ shuffle) at both Reynolds numbers.  At $\text{Re}_\lambda \approx 1300$, center $c = 11$ (top panels): the dissipation-conditioned quiescent stratum drops to the Markov floor at $s^* = 8$ ($\Delta r = 1.26$), while the intermittent stratum persists to $s^* = 20$ ($\Delta r = 3.15$).  The increment-amplitude stratification confirms: the core gives $\Delta r = 1.10$, the tails give $\Delta r = 3.31$.  At this center, the quiescent/core component recovers the canonical $\Delta r \approx 1.1$--$1.3$, and the excess memory resides entirely in intermittent events.  The stratification is scale-dependent: at $c = 14$ ($\ell_2/\lambda_T = 2.02$), the tail still carries more memory than the core ($\Delta r_\mathrm{core} = 2.68$, $\Delta r_\mathrm{tail} = 3.94$), but both exceed the canonical value.  At $c = 8$ (near dissipation, $\ell_2 = 0.78\lambda_T$), neither stratum approaches $\Delta r \approx 1$ ($\Delta r_\mathrm{quiescent} = 3.62$, $\Delta r_\mathrm{intermittent} = 3.78$).  The quiescent recovery to $\Delta r \approx 1$ is most pronounced at the mid-inertial-range center $c = 11$ and weakens at scales closer to the dissipation or forcing ranges (see Supplemental Material~\cite{SM}).

Near the dissipation range ($c = 8$, $\ell_2 = 0.78\lambda_T$), the increment-amplitude stratification reverses: the core carries more memory ($\Delta r = 3.78$) than the tails ($\Delta r = 2.36$).  This is consistent with viscous coupling dominating the conditional structure at near-dissipation scales, where spatial correlations induced by the viscous term affect all increment amplitudes, while the rare extreme events that characterize inertial-range intermittency are less prominent.

A forcing-edge cap analysis (restricting $i_3 \leq 24$, i.e., $\ell_3 \leq 9.7\lambda_T$; see End Matter) at $c = 11$ yields $\Delta r = 3.15$---identical to the uncapped value.  The excess memory is internal to the inertial range, not an artifact of forcing or integral-scale contamination.

At $\text{Re}_\lambda \approx 433$ (bottom panels of Fig.~\ref{fig:stratified}), center $c = 8$ ($\ell_2/\lambda_T = 0.93$), the dissipation-conditioned quiescent stratum gives $\Delta r = 1.36$ and the core gives $\Delta r = 1.36$. Both match the canonical value and the $\text{Re}_\lambda \approx 1300$ result at $c = 11$.  The intermittent stratum and tails exceed the 15-scale grid's maximum reach of $\Delta r = 2.72$.  The excess memory is present but not fully resolvable at this Re due to the narrower inertial range.  Three of four full-cascade centers at $\text{Re}_\lambda \approx 433$ are unresolved ($\Delta r > 2.72$), confirming that the full-cascade memory exceeds the canonical value at this Re as well.

\begin{figure}[htb]
\includegraphics[width=0.7\columnwidth]{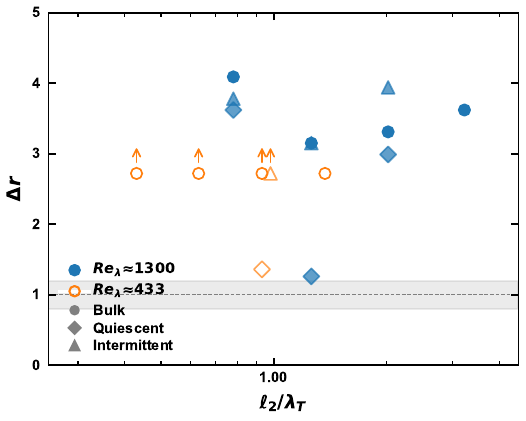}
\caption{Summary: $\Delta r$ vs.\ $\ell_2/\lambda_T$ at both Re.  Filled symbols: $\text{Re}_\lambda \approx 1300$ (30-scale grid).  Open symbols: $\text{Re}_\lambda \approx 433$ (15-scale grid).  Circles: full cascade.  Diamonds: dissipation-conditioned quiescent stratum.  Triangles: dissipation-conditioned intermittent stratum.  Upward arrows: lower bounds ($\Delta r$ exceeds grid reach).  Horizontal dashed line: canonical $\Delta r = 1$~\cite{Stresing2011}.  Gray band: $\Delta r = 0.8$--$1.2$ (Stresing et al.\ uncertainty range).  The quiescent recovery to $\Delta r \approx 1$ is most pronounced at mid-inertial-range centers ($\ell_2/\lambda_T \approx 1$); at other centers the quiescent stratum also carries elevated memory.}
\label{fig:summary}
\end{figure}

Figure~\ref{fig:summary} summarizes the measured $\Delta r$ across both Reynolds numbers.  At the mid-inertial-range centers ($c = 11$ at $\text{Re}_\lambda \approx 1300$; $c = 8$ at $\text{Re}_\lambda \approx 433$), the quiescent values (diamonds) fall within the Stresing et al.~\cite{Stresing2011} consensus band $\Delta r \approx 1$.  At centers closer to the dissipation range or deeper into the inertial range, even the quiescent stratum carries elevated memory ($\Delta r_\mathrm{quiescent} = 3.0$--$3.6$).  The intermittent/tail values (triangles) sit at $\Delta r \approx 3$--$4$ at $\text{Re}_\lambda \approx 1300$ and exceed $\Delta r = 2.7$ at $\text{Re}_\lambda \approx 433$.  The full-cascade values (circles) are dominated by the intermittent contribution.

\emph{Discussion.}---Five factors distinguish our results from the current consensus: (a)~statistical power (${\sim}10^6$ vs ${\sim}10^5$ samples per scale); (b)~DNS data eliminating probe noise and the Taylor hypothesis; (c)~two independent surrogate baselines pinning the Markov floor; (d)~higher $\text{Re}_\lambda$ in the primary dataset; and (e)~stratification revealing the quiescent and intermittent components separately.

The cross-check at $\text{Re}_\lambda \approx 433$ partly disentangles factor~(d) from factors~(a)--(c).  Even at a Reynolds number within the experimental range of the original measurements, the intermittency-driven excess memory appears when the statistical power and surrogate methodology are present.  This suggests that factor~(a)---statistical power---is the primary reason the excess was not detected previously.  The prior experimental consensus was likely measuring the quiescent component of the cascade at mid-inertial-range scales, which our stratification shows satisfies $\Delta r \approx 1$ at the corresponding centers in both datasets.

The elevated $\Delta r$ at $c = 8$ ($\ell_2 = 0.78\lambda_T$), where the reversed stratification pattern is observed, coincides with the spectral bottleneck region~\cite{Sinhuber2017}.  The bottleneck introduces non-local correlations that plausibly inflate the apparent coherence length near the dissipation range.  However, the forcing-edge cap analysis at $c = 11$ rules out propagation from the bottleneck region as the source of inertial-range memory, and the amplitude-dependent stratification at $c = 11$ (quiescent $\Delta r \approx 1$, intermittent $\Delta r \approx 3$) is inconsistent with a spectral (amplitude-independent) mechanism.  The $\text{Re}_\lambda \approx 433$ cross-check provides additional evidence: the bottleneck amplitude decreases with $\text{Re}_\lambda$~\cite{Sinhuber2017}, yet the stratification pattern is essentially unchanged.

The implications for the cascade Fokker--Planck framework~\cite{Friedrich1997,Friedrich2011} and for applications built on it are significant but not catastrophic.  The Markov approximation requires $\Delta r \gtrsim 3$ for the full cascade, substantially more restrictive than previously assumed.  However, at mid-inertial-range scales, the quiescent cascade, representing half the samples by construction (median split), remains Markov at the classical threshold.  This suggests a two-component description: a Markov FP equation for the quiescent cascade plus an intermittency-dependent memory correction, possibly via a generalized Langevin equation with a scale-dependent kernel.

For the IFT of the cascade~\cite{Nickelsen2013,Reinke2018,Fuchs2020}, which assumes the Markov property, whether the theorem remains approximately valid is an open question.  The intermittent events that break Markovianity at $\Delta r < 3$ may or may not contribute negligibly to the entropy-production average that the IFT constrains.  A direct test, stratifying the IFT verification by dissipation intensity, is needed.  If a non-Markovian correction is required, the Speck--Seifert generalization~\cite{Speck2007} provides the appropriate framework.

The quantitative consistency of the quiescent $\Delta r \approx 1.1$--$1.4$ at the mid-inertial-range centers across a factor of~3 in $\text{Re}_\lambda$ ($c = 11$ at $\text{Re}_\lambda \approx 1300$, $c = 8$ at $\text{Re}_\lambda \approx 433$) is notable.  It suggests the canonical $\lambda_\mathrm{EM} \approx \lambda_T$ reflects a genuine physical property of the quiescent cascade at scales $\ell_2 \sim \lambda_T$, robustly measurable across Reynolds numbers, between DNS and experiment, and independent of the specific nesting structure~\cite{Stresing2011}.  That this recovery weakens at other centers, e.g., deeper in the inertial range ($c = 14$, $\Delta r_\mathrm{quiescent} \approx 3.0$) or near dissipation ($c = 8$, $\Delta r_\mathrm{quiescent} \approx 3.6$), suggests the quiescent Markov property is itself scale-dependent, not merely an amplitude-independent spectral feature.

\emph{Conclusions.}---The Markov--Einstein coherence length of the full turbulent cascade is determined to be $\Delta r \approx 3.2$--$3.6$ at $\text{Re}_\lambda \approx 1300$, approximately $3\times$ the canonical value.  The excess is driven by intermittent events: at mid-inertial-range scales ($\ell_2 \sim \lambda_T$), the quiescent cascade recovers $\Delta r \approx 1$ while intermittent events require $\Delta r \approx 3$--$4$.  This recovery is itself scale-dependent, weakening at centers deeper in the inertial range.  Near the dissipation range, the mechanism reverses, consistent with the spectral bottleneck.  The memory is internal to the inertial range, and the finding is Reynolds-number-independent over $\text{Re}_\lambda \approx 433$--$1300$.  A non-Markovian description is empirically required for the intermittent component; the Markov FP framework remains valid for the quiescent cascade at mid-inertial-range scales.

\bibliography{references}

% === End Matter ===

\appendix*

\emph{$\Delta r \leftrightarrow$ physical-scale conversion.}---The cascade coordinate is $r = -\ln(\ell/L)$.  A log-scale gap $\Delta r$ corresponds to a physical scale ratio: $\ell_1/\ell_2 = \exp(-\Delta r)$.  At the center scale $\ell_2 = \lambda_T$ and with the canonical physical coherence length $\lambda_\mathrm{EM} \approx 0.8\lambda_T$~\cite{Stresing2011}, the corresponding log-scale gap is $\Delta r = |\ln[(\lambda_T + 0.8\lambda_T)/(\lambda_T - 0.8\lambda_T)]| = \ln(1.8/0.2) = \ln 9 \approx 2.2$.  This overstates the gap because the increment nesting structure means the ``step'' is not centered symmetrically.  In the actual gap-scan parameterization used by Stresing et al.~\cite{Stresing2011} and in this work, the effective $\Delta r$ at $\ell_2 \sim \lambda_T$ for a physical gap of ${\sim}\lambda_T$ corresponds to $\Delta r \approx 1$, verified by comparing the gap index~$s^*$ with the physical $\ell$-separation at each center.  Our measurements and the literature values are compared consistently in the same $\Delta r$ units throughout.

\emph{Forcing-edge cap analysis.}---At center $c = 11$ ($\ell_2/\lambda_T = 1.26$) in the $\text{Re}_\lambda \approx 1300$ dataset, we repeated the asymmetric gap scan with the restriction $i_3 \leq 24$, corresponding to $\ell_3 \leq 9.7\lambda_T$.  This caps the ``past'' scale well below the integral scale $L_1 \approx 70\lambda_T$, testing whether the excess memory originates from energy-containing or forcing-range scales.  The result: $\Delta r = 3.15$, identical to the uncapped measurement.  At centers $c = 14$ and $c = 17$, the cap pushes $i_1$ into the dissipation range at large separations, preventing clean comparison.  The $c = 11$ test is the cleanest available, and its result rules out large-scale contamination as the source of the excess memory.

\end{document}